\documentclass[aps,twocolumn,pra,showpacs]{revtex4-1}
\usepackage{graphicx}
\usepackage[dvipdfm]{hyperref}
\usepackage[normalem]{ulem}
\usepackage{amssymb}

\begin{document}

\title{Zero-temperature phase diagram of Yukawa bosons}
\author{O. N. Osychenko}
\author{G. E. Astrakharchik}
\author{F. Mazzanti}
\author{J. Boronat}
\affiliation{Departament de F\'{\i}sica i Enginyeria Nuclear,  Universitat Polit\`ecnica de Catalunya, Campus Nord B4-B5, 08034 Barcelona, Spain}
%http://www.aip.org/pacs
%\pacs{64.60.-i}%{General studies of phase transitions}
%\pacs{03.75.Ss}%{Degenerate Fermi gases}
%\pacs{71.10.Hf}%{Non-Fermi-liquid ground states, electron phase diagrams and phase transitions in model systems}
%03.75.Ss	Degenerate Fermi gases
%05.30.Fk	Fermion systems and electron gas (see also 71.10.-w Theories and models of many-electron systems; see also 67.10.Db Fermion degeneracy in quantum fluids)
%03.75.Hh	Static properties of condensates; thermodynamical, statistical, and structural properties
%64.60.-i	General studies of phase transitions
%71.10.Hf	Non-Fermi-liquid ground states, electron phase diagrams and phase transitions in model systems
\date{\today}

\begin{abstract}
We study the zero-temperature phase diagram of bosons interacting via screened Coulomb (Yukawa) potential by means of the diffusion Monte Carlo method. The Yukawa potential is used as a model interaction in the neutron matter, dusty plasmas and charged colloids. As shown by D. S. Petrov {et al.} [Phys. Rev. Lett. {\bf 99}, 130407 (2007)], interactions between weakly bound molecules of heavy and light fermionic atoms are described by an effective Yukawa potential with a strength related to the heavy-light mass ratio $M/m$ which might lead to crystallization in a two-dimensional geometry if the mass ratio of heavy-light fermions exceeds a certain critical value. In the present work we do a thorough study of the quantum three-dimensional Yukawa system. For strong interactions (equivalently, large mass ratios) the system experiences several phase transitions as the density is increased, passing from gas to solid and to gas phase again. Weakly interacting Yukawa particles do not crystallize at any density. We find the minimal interaction strength at which the crystallization happens. In terms of the two-component fermionic system, this strength corresponds to a heavy-light mass ratio of $M/m\sim 180$, so that it is impossible to realize the gas-crystal transition in a conventional bulk system. For the Yukawa model of fermionic mixtures we also analyze the possibility of building molecular systems with very large effective mass ratios by confining the heavy component to a sufficiently deep optical lattice. We show how the effective mass of the heavy component can be made arbitrarily large by increasing the lattice depth, thus leading to a tunable effective mass ratio that can be used to realize a molecular superlattice.
\end{abstract}

\maketitle

\section{Introduction}

Recent advances in trapping and controlling ultracold dilute gases have permitted the realization of highly tunable and extremely pure Fermi systems \cite{RPMcoldgases}. This has provided new insight in the study of fundamental problems in condensed matter physics. For example, the original BCS theory\cite{BCS} was developed to explain superconductivity in metals, where the control over interactions and densities is very limited. However, in recent experiments with ultracold Fermi gases in the BCS-BEC crossover the strength of the interactions is controlled by external magnetic fields in the vicinity of a Feshbach resonance, while the geometry is tuned by means of magnetic or optical confinement. This has allowed, for instance, the measurement of the equation of state in the BCS-BEC crossover in high precision experiments\cite{EOS:experiment}. Numerically, the best calculation of the zero-temperature equation of state is obtained in quantum Monte Carlo simulations\cite{EOS:BCSBEC3D}

After the big success achieved with single species there is nowadays a growing interest in fermionic mixtures. Quite recently, fermionic mixtures consisting of atoms with different masses have been realized experimentally~\cite{exp:LiK,exp:LiYb} and studied theoretically~\cite{th:LiK}. Novel physical phenomena such as Efimov states\cite{theory:Efimov}, trimer and cluster formation might be observed\cite{exp:Efimov} in these systems. The case of large mass imbalance is especially interesting, and mixtures of $^6$Li and $^{40}$K are being investigated experimentally\cite{exp:LiK}. Even larger mass ratios are reached in mixtures of $^6$Li and $^{173}$Yb\cite{exp:LiYb}. In this article we present results for the phase diagram of Fermi mixtures as a function of the mass ratio using quantum Monte Carlo methods and determine how crystallization of this system can be realized.

From the theoretical point of view, it was proposed in Ref.~\cite{YukawaPot2D} that an effective Yukawa interaction, induced between heavy-light pairs of fermions, might lead to crystallization in quasi-two-dimensional systems. In this work we extend that discussion and analyze the possibility of realizing a gas-crystal phase transition at zero temperature in three-dimensional (3D) systems. We obtain the phase diagram and discuss how large mass ratios have to be for reaching crystallization.

The interest in the phase diagram of quantum Yukawa particles is rather old as the Yukawa potential has long been used, for instance, as a model for neutron matter~\cite{neutron}.
The Yukawa potential also describes interactions in dusty plasmas where charged dust particles are surrounded by plasma which introduces screening \cite{dusty plasma}. The Yukawa potential is often used as well as a model for suspensions of charged colloidal particles~\cite{yukawa:colloids}. The classical finite temperature phase diagram has been extensively studied \cite{dusty plasma,yukawa:colloids} while much less is known about the full quantum phase diagram.

In the 70's, Ceperley and collaborators~\cite{Ceperley} used the diffusion Monte Carlo algorithm to estimate the zero-temperature phase diagram of the Yukawa Bose fluid. In their work the phase diagram was built assuming that the Lindemann ratio remains constant along the solid-gas coexistence curve, with the explicit value being evaluated only at a single point. In the present work we carry out a full study of the transition curve and present the phase diagram in terms of experimentally relevant densities and mass ratios of heavy to light fermions. The Lindemann criterion prediction has turned out to be quite precise apart from the region of high densities.

\section{Hamiltonian}

Mixtures of fermions with different masses have been realized recently in a new generation of experiments \cite{exp:LiK,exp:LiYb}. The interactions can be tuned to allow the formation of two-component molecules. The $s$-wave interactions within a single component are prohibited due to the Pauli principle. Yet, an effective interaction between same-spin fermions can be induced by the presence of the other component. The limit of large mass ratio has been analytically addressed in Ref.~\cite{YukawaPot2D}. The effective interaction between heavy particles, which was obtained in the limit of large distances within first Born approximation, has the form of a screened Coulomb (Yukawa) potential. This leads to a description of the system in terms of a composite (molecular) bosonic gas interacting with an effective potential. The effective $p$-wave interaction between heavy particles in BEC-BCS crossover, derived in Ref.~\cite{Nishida}, is no more of Yukawa form but rather present oscillations that increase in amplitude when going from BCS to BEC regime. In the present article we limit ourselves to considering the $s$-symmetry effective Yukawa interaction most relevant on the BEC side of the BEC-BCS crossover.

We study a system of heavy fermions of mass $M$ interacting among themselves and moving on a background of light fermions of mass $m$. The net effect induced by the movement of the light fermions can be characterized by a Yukawa potential, leading to the following effective  Hamiltonian~\cite{YukawaPot2D} describing the interaction between composite bosons formed by pairs of heavy and light atoms
\begin{equation}
\hat H=-\frac{\hbar^2}{2M}\sum_i\Delta_{i} + \sum_{i<j}\frac{2\hbar^2}{m}\frac{\exp(-2|{\bf r}_i-{\bf r}_j|/a)}
{a|{\bf r}_i-{\bf r}_j|},
\label{H}
\end{equation}
where $a$ is the atom-atom $s$-wave scattering length and ${\bf r}_i$ are positions of heavy atoms while the positions of light atoms have been integrated out. The ground-state properties of the system are then governed by two dimensionless parameters, namely the gas parameter $na^3$ and the mass ratio $M/m$. Equivalently, Hamiltonian (\ref{H}) describes a bosonic system interacting via the screened Coulomb potential $V_{int}(r) = q \exp(-\lambda r)/r$ by mapping the charge to $q = 2\hbar^2/ma$ and the screening length to $\lambda^{-1} = a/2$.

We calculate the ground-state properties corresponding to the Hamiltonian~(\ref{H}) by means of the diffusion Monte Carlo (DMC) algorithm~\cite{DMC}. This method solves stochastically the Schr\"odinger equation in imaginary time providing the exact energy within controllable statistical errors. The coexistence curves can then be traced by direct comparison of the energies of the solid and gas phases. The efficiency of the DMC method is greatly enhanced when importance sampling is used. This is done by multiplying the (unknown) ground-state wave function $\psi({\bf r}_1, \ldots, {\bf r}_N)$ by a guiding wave function $\psi_T({\bf r}_1, \ldots, {\bf r}_N)$ and solving the equivalent Schr\"odinger equation for the product. As a result, the points in phase space where the guiding function is large get sampled more frequently and this improves convergence to the ground state.

The properties of the gas phase are studied by constructing the guiding function in a Bijl-Jastrow two-body product form $\psi_T({\bf   r}_1, \ldots, {\bf r}_N) = \prod_{i<j}f_2(|{\bf r}_i-{\bf r}_j|)$. We determine the optimal two-body Jastrow term $f_2(r)$ by solving the corresponding Euler--Lagrange hypernetted-chain equations\cite{HNC} (HNC/EL) discarding the contribution of the elementary diagrams. In this scheme the static structure factor $S(k)$ that minimizes the variational energy in the subspace of Jastrow wave functions has the form
\begin{equation}
S(k) = { t(k) \over \sqrt{ t^2(k) + 2 t(k) V_{ph}(k)}} \ ,
\label{hncel-b1}
\end{equation}
with $t(k)=\hbar^2k^2/2m$ and $V_{ph}(k)$ the so-called particle-hole interaction.
Its Fourier transform $FT[V_{ph}(k)] = \tilde V_{ph}(r)$ satisfies the following equation in coordinate space
\begin{equation}
\tilde V_{ph}(r) = g(r) V(r) + {\hbar^2 \over m}\!\mid\!\nabla\!\sqrt{g(r)}\!\mid^2\!
+ (g(r)\!-\!1) \omega_I(r) \ ,
\label{hncel-b2}
\end{equation}
where $V(r)$ and $g(r)$ are the bare two-body potential and the pair distribution function (the Fourier transform of $S(k)$), respectively. Finally, in momentum space the induced interaction $\omega_I(k)$ becomes
\begin{equation}
\omega_I(k) = -{1\over 2}\,t(k) { [ 2 S(k)+1 ][ Sk)-1 ]^2 \over S^2(k) } \ .
\label{hncel-b3}
\end{equation}
In this way, Eqs.~(\ref{hncel-b1}),~(\ref{hncel-b2}) and~(\ref{hncel-b3}) form a set of nonlinear coupled equations that have to be solved   iteratively. The Fourier transform of the resulting $S(k)$ provides $g(r)$ and, in this scheme, the optimal two-body Jastrow factor results from the corresponding HNC/0 equation
\begin{equation}
%g(r) = [f_2(r)]^2\,e^{N(r)} \ ,
f_2(r) = \sqrt{g(r)}e^{-N(r)/2}\ ,
\end{equation}
where $N(r)$ is the sum of nodal diagrams, related to $S(k)$ in momentum space by the expression $N(k)=[S(k)-1]^2/S(k)$.

The resulting wave function captures basic ingredients coming both from the two- and many-body physics of the problem. On the other hand, the energy of the solid phase is obtained by using a Nosanow-Jastrow guiding wave function $\psi_T({\bf r}_1, \ldots, {\bf r}_N) = \prod_{i=1}^N f_1({\bf r}_i-{\bf r}_i^{\text{latt.}})\prod_{i<j}f_2(|{\bf r}_i-{\bf r}_j|)$ with Gaussian one-body terms $f_1({\bf r}_i-{\bf r}_i^{\text{latt.}}) = \exp(-\alpha({\bf r}-{\bf r}_i^{\text{latt.}})^2)$ describing the localization of particles close to the lattice sites ${\bf r}_i^{\text{latt.}}$. The parameter $\alpha$ controls the localization strength and is optimized by minimizing the variational energy.

In order to find the energy in the thermodynamic limit, we carry out simulations of a system of $N$ particles in a box with periodic boundary conditions, and take the limit $N\to\infty$ while keeping the density fixed. In the simulation of the crystal the number of particles should be commensurate with the box which restricts the allowed number of particles. For FCC packing the simulation box supports
$N=4i^3=4, 32, 108, 256, \ldots$. In order to add more values we also use periodic boundary conditions on a truncated octahedron, which allows simulations with $N=2i^3 = 2, 16, 54, 128, 250, 432, \ldots$ particles with a larger effective volume of the simulation box and reduced anisotropy effects. Finally, the convergence is further improved by the Ewald summation technique\cite{Ewald} which we use in the calculations at large densities.

\begin{figure}
\includegraphics[width=0.8\columnwidth, angle=-90]{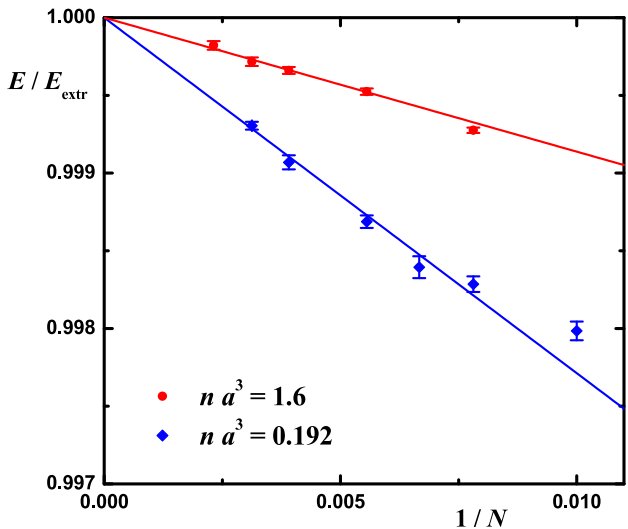}
\caption{(Color online) An example of finite-size dependence of the energy in the gas with periodic boundary conditions in truncated octahedron for $M/m=187$ at two different densities $na^3 = 1.6$ (upper set of data points) and $na^3 = 0.192$ (lower set of data points). Symbols, DMC energy; lines, linear fit to energy for large system sizes.
Energies are scaled with the thermodynamic value $E_{extr}$, obtained in $1/N\to 0$ extrapolation.}
\label{Fig:finite size}.
\end{figure}

In Fig.~\ref{Fig:finite size} we show two characteristic examples of the finite-size dependence of the energy at two different densities.
For large enough system sizes, the energy is well fitted by a linear dependence in $1/N$. For small number of particles the behavior is no longer linear, especially at large densities due to strong interparticle correlations. We find that system sizes of $N>100$ have to be used in order to ensure the linear regime at considered densities. The thermodynamic energy is then obtained as a result of a linear extrapolation $1/N \to 0$.

\section{Phase diagram}

An intrinsic property of Coulomb particles is to self assemble into a Wigner crystal at low densities and to remain in a gaseous phase in the opposite limit, due to the long-range character of the interaction~\cite{Wigner}. The Yukawa potential is similar to the Coulomb one at densities large enough for the interparticle distance to be much smaller than the screening length, which is fixed by the $s$-wave scattering length $a$. One then concludes that the Yukawa system stays in a gaseous phase at large densities. In the opposite regime of small densities, $na^3\ll 1$, the interaction potential decays exponentially fast showing a short-range behavior that leads the system to a gaseous phase. For example, the FCC crystal of hard-sphere bosons of diameter $a_s$ melts at density $n a_s^3 \approx 0.24$ \cite{HS3D}. The intermediate regime $na^3\approx 1$ is the most interesting one, as crystallization may or may not take place depending on the strength of the interaction, which in the current case of the Hamiltonian in Eq.~(\ref{H}) is governed by the mass ratio $M/m$. A relevant question then is what is the minimal mass ratio at which crystallization can be observed.

\begin{figure}
%\begin{center}
\includegraphics[width=0.8\columnwidth, angle=-90]{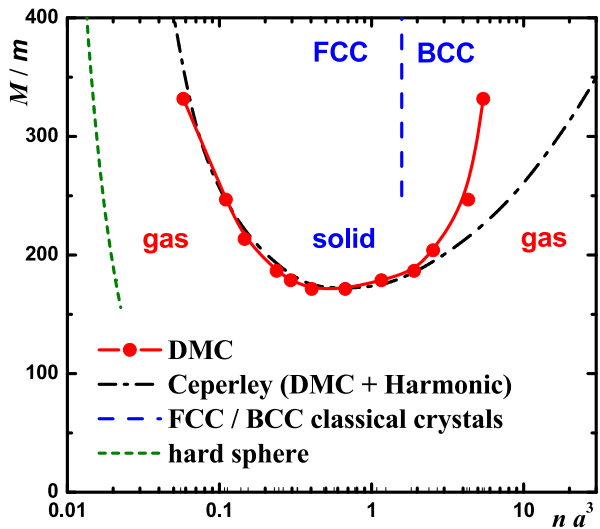}
\caption{(Color online) Zero-temperature phase diagram of the Yukawa potential corresponding to the Hamiltonian in Eq.~(\ref{H}) in terms of the gas parameter $na^3$ and the mass ratio $M/m$. Red symbols: transition point as obtained from the Maxwell double-tangent construction applied to the Monte Carlo data energies extrapolated to the thermodynamic limit; dashed line: critical density $n a^3 = 1.58\ldots$ at which the energy of perfect FCC and BCC packings are equal; dash-dotted line: prediction of Ceperley {et al.}~\cite{Ceperley} obtained by imposing a constant Lindemann ratio; short-dashed line: $na_s^3=0.24$\cite{HS3D}.}
\label{Fig:phase diagram}.
%\end{center}
\end{figure}

In order to obtain an accurate description of the phase diagram, we study the finite size dependence and extrapolate the energy to the thermodynamic limit. The resulting energies of the gas and solid phases are then analyzed using the double-tangent Maxwell construction which provides the melting and freezing densities. The zero-temperature phase diagram parametrized in terms of the dimensionless density $na^3$ and the mass ratio, is shown in Fig.~\ref{Fig:phase diagram}. We find that for mass ratios smaller than the critical value $M/m\approx 180$ the gas phase is energetically preferable at any density. On the other hand, for larger mass ratios there is always a gas-solid transition at low densities and a solid-gas transition at large ones. Energetically, both the FCC and BCC lattices are possible in the solid phase. It is very difficult to discern numerically which packing is preferred as the energies in different crystalline phases are extremely close. Still, in the large potential energy limit, corresponding to a mass ratio $M/m\gg 1$, it is enough to compare the potential energy of the classical crystals with different packings. A simple, geometrical construction assuming that particles are tightly tied to their equilibrium positions leads to a transition density $n a^3 \approx 1.58$. This prediction is depicted as a blue dashed line in Fig.~\ref{Fig:phase diagram}. In the low-density limit we numerically find the value of the $s$-wave scattering length $a_s$ of the Yukawa potential~(\ref{H}) and fit it as $a_s/a = 0.436 \ln(M/m)$ with accuracy below 1\% in the region of interest. Note that $a$ is the $s$-wave scattering length of fermionic particles which lead to the effective bosonic Hamiltonian~(\ref{H}) while $a_s$ is the $s$-wave scattering length between bosonic Yukawa particles. For the sake of comparison we also plot in Fig.~\ref{Fig:phase diagram} the gas-solid transition line of hard spheres of size $a_s$ given by $M/m = \exp(1.424/(n^{1/3}a))$.

The figure also shows the results of Ceperley {et al.}~\cite{Ceperley} which were obtained by doing DMC calculations for three characteristic points in the phase diagram close to the solid-gas transition line. Overall, the agreement between that prediction and our results is good, the main differences affecting the region of large density where Coulomb effects are strong. To our best knowledge this is the first time that the high-density quantum solid-gas phase transition is observed in a simulation of Yukawa systems.

In the case of the fermionic molecules, the resulting critical mass ratio is much larger than $M/m\approx 13.6$ for which the system is unstable due to formation of Efimov states\cite{theory:Efimov}. The obtained phase diagram describes properties of metastable fermionic molecules while the true ground state corresponds to a many-body bound state. The stronger the effective interaction is (that is, the larger the mass ratio), the more distant are heavy fermions and the smaller the overlap with localized Efimov states is.

\section{Reaching large mass ratios}

According to our results, the minimal mass ratio for which the crystalline phase can exist is $M/m\approx 180$ and it is achieved at the somewhat large value of the gas parameter $na^3\approx 0.3$. At these densities the fermionic nature of the molecules becomes important as the Hamiltonian~(\ref{H}) is derived under the assumption that $na^3\lesssim 1/8$\cite{YukawaPot2D}. Our bosonic model is expected to be reliable at smaller densities where the critical mass ratio is further increased.

The mixtures of different fermionic atoms have already been successfully realized in experiments\cite{exp:LiYb} but at significantly smaller mass ratios. Probably, the largest directly achievable mass ratio currently is that of Yb and Li atoms, $M/m = 29$, which is still much smaller than the critical mass ratio needed to observe the formation of an ultracold crystal.

An alternative way to realize a fermionic mixture with a large and variable mass ratio is to confine one of the components to an optical lattice. At low filling fraction the distances between atoms are large compared with the lattice spacing, and the separation of length scales allows the description of the movement of a particle in the lattice as that of a quasiparticle with an effective mass moving in a medium where the lattice is absent. In a deep lattice interactions between particles are much weaker than the confining energy and so, to a first approximation, one can consider that as the problem of a single particle diffusing in the lattice.

An optical lattice created by counter-propagating laser beams imposes an external potential  $V_{\text{latt.}}(x,y,z)=V_{0}\left(\sin^2{kx}+\sin^2{ky}+\sin^2{kz}\right)$ on every particle. The diffusion of a particle over a large distance is then governed by the tunneling rate between neighboring sites. The diffusion is largely suppressed (and the effective mass greatly increased) when the amplitude of the optical lattice is large, i.e. when $V_0\gg E_r$ with $E_r=\hbar^2k^2/2m$ the recoil energy. The excitation spectrum in the lowest band can be described by Bloch waves of quasi-momentum $\mathbf{q}$ and energy $\varepsilon_0(\mathbf{q})=\frac{3}{2}\hbar\omega_0 - 2J\left(\cos{q_xd}+\cos{q_yd} +\cos{q_zd}\right) +\ldots$ with $d=\pi/k$ the lattice constant~\cite{revmodBloch}. At small momenta the spectrum is quadratic in $q$ and can be interpreted as the spectrum $\varepsilon_0(q) = E_0 + \hbar^2q^2/2m^*$ of a free quasiparticle with an effective mass $m^*$. Within the lowest band approximation the effective mass is inversely proportional to the hopping parameter $J$,
\begin{equation}
\frac{m^*}{m} = \frac{1}{\pi^2}\frac{E_r}{J} \ .
\label{m:band}
\end{equation}
The tunneling is greatly suppressed in the deep optical lattice limit $V_0\gg E_r$. To better understand the contribution of the tunneling term in the present case, a semiclassical treatment within the Wentzel-Kramers-Brillouin(WKB) approximation can be used to calculate the tunneling probability. One finds that it is proportional to $J^2 \propto \exp\{-2\int_{x_1}^{x_2}\;dx\,\sqrt{2m(V(x)-E)}/\hbar\}$, where $x_1$ and $x_2$ are the classical turning points. In the deep optical lattice limit the energy of the moving particle is only slightly larger than the potential energy at a lattice site, and therefore $V(r) - E \approx V(r)$, so $x_1$ and $x_2$ correspond to the positions of two neighboring minima. The resulting integral can be easily evaluated and predicts an exponential form $J \propto \exp(-\sqrt{V_0/E_r})$. A more precise expression can be obtained from the width of the lowest band in the 1D Mathieu-equation~\cite{revmodBloch}, yielding
\begin{equation}
%see Abramowitz Stegun 20.2.31)
J=\frac{4}{\sqrt\pi}E_r\left(\frac{V_0}{E_r}\right)^{3/4}
\exp\left\{-2\left(\frac{V_0}{E_r}\right)^{1/2}\right\} \, .
\label{J:band}
\end{equation}
This expression, together with Eq.~(\ref{m:band}), provides an analytic approximation for the effective mass $m^*$.

\begin{figure}
\begin{center}
\includegraphics[width=0.8\columnwidth, angle=-90]{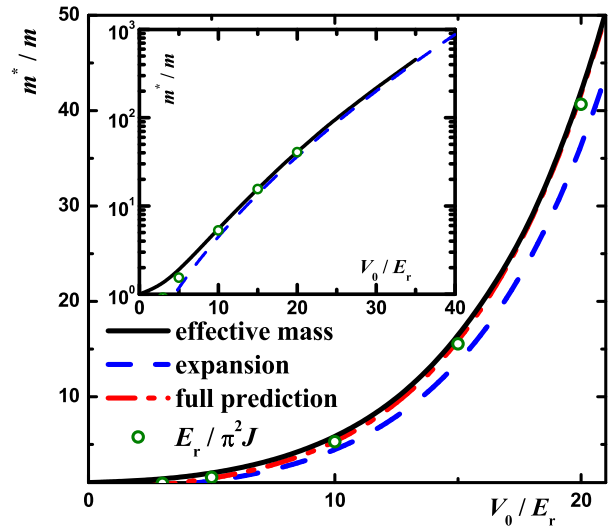}
\caption{(Color online) Effective mass as a function of the lattice amplitude $V_0$ in units of the recoil energy $E_r$. Solid line: results obtained from the diffusion constant evaluated by propagation in imaginary-time; circles: lowest band approximation of Eq.~(\ref{m:band}) with values of $J$ taken from~\cite{revmodBloch}; dashed line: same results with $J$ from the expansion in Eq.~(\ref{J:band}); dash-dotted line, same expansion with $V_0$ shifted by $-3/4 E_r$. Inset: same results on a semi-logarithmic scale.}
\label{Fig:meff}
\end{center}
\end{figure}

In order to determine the dependence of $m^*$ on the lattice parameters in a non-perturbative way we evaluate the diffusion constant $D$ of a real particle moving on the lattice and compare it to the diffusion constant $D_0 = \hbar^2/2m^*$ of a free quasiparticle of effective mass $m^*$. The diffusion constant is obtained by means of DMC propagation in imaginary time by measuring the mean-square displacement
$\langle({\bf r}(\tau) - {\bf r}(0))^2 \rangle = \langle(x(\tau) - x(0))^2 \rangle+\langle(y(\tau)-y(0))^2 \rangle+\langle(z(\tau)-z(0))^2 \rangle$ where ${\bf r}=(x,y,z)$ denote particle coordinates. The diffusion constant is then extracted as $D = \lim\limits_{\tau \to \infty} \hbar \langle({\bf r}(\tau) - {\bf r}(0))^2 \rangle/(2\tau d)$, where $d=3$ is the system dimensionality. The resulting dependence of $m^*$ on the lattice amplitude is shown in Fig.~\ref{Fig:meff}. The figure shows the Monte Carlo prediction (solid line) compared with the approximation of Eq.~(\ref{m:band}) with $J$ taken from  Ref.~\cite{revmodBloch} (circles) and from Eq.~(\ref{J:band}) (dashed line). As it can be seen, there is an almost constant shift between $m^*$ obtained in the Monte Carlo simulation and Ref.~\cite{revmodBloch} compared to Eqs.~(\ref{m:band}-\ref{J:band}). We have found that the description in the relevant region of interest is very much improved by subtracting a constant shift $E^{(1)}=-3/4 E_r$ from $V_0$ in the argument of Eq.~(\ref{J:band}). This last prediction is shown by a thin line in Fig.~\ref{Fig:meff} and provides a good approximation for $V_0\gtrsim 10 E_r$.

One can understand these results in the following way: in the absence of the optical lattice the effective mass and the bare mass coincide, so $m^* = m$. As the amplitude $V_0$ of the lattice is increased, the particle movement is slowed down and the effective mass increases. In the deep optical lattice limit the effective mass grows as $m^*/m\propto\exp(\sqrt{V_0/E_r})$ and so the ratio can be made arbitrarily large by increasing the amplitude $V_0$ (for instance $m^*/m \sim 1000$ at $V_0/E_r = 40$; see the inset in Fig.~\ref{Fig:meff}). This mechanism allows for increasing the mass of one of the two components while keeping the other one unaltered, so that the ratio $M/m$ of the fermionic mixture can be made as large as desired when the mass of the heavy component is identified with the effective mass $m^*$. Consequently, and according to the phase diagram shown in Fig.~\ref{Fig:phase diagram}, there is a wide range of densities where one could find the system in the crystalline superlattice phase. Heights of optical lattices as large as $(35-60) E_r$ are readily achieved in current experiments \cite{large_recoil} and correspond to sufficiently large effective mass ratios for the crystallization to be realized.

Both small density and large density transition lines are accessible for Yukawa interaction caused by screening in dusty plasma, colloids and neutron matter. On the contrary, in two-component Fermi gas only the left part of the phase diagram can be realized since the effective Yukawa interaction is valid only at low densities. In fact, the validity criterion for the interaction potential in Eq.~(\ref{H}) was studied in Ref.~\cite{YukawaPot2D} and was found to be well satisfied for distances larger then $r \approx 2a$ which leads to the condition $\rho a^3 \lesssim 1/8$ when $r$ is identified with the mean interparticle distance. In this way, for example, for $\rho a^3 = 0.1$ and mass ratio $M/m = 300$ the system is expected to be in a crystalline form. Much larger effective mass ratios can be achieved for realistic\cite{large_recoil} lattice heights of $(35-60) E_r$. We thus conclude that by using an optical lattice, a fermionic mixture of very different mass components can be used to test the phase diagram of the equivalent Yukawa model.

\section{Summary and conclusions}

To summarize, in this work we have obtained the zero-temperature phase diagram of bosons interacting through Yukawa forces. We have used a diffusion Monte Carlo simulation starting from a very good approximation to the optimal variational ground-state wave function obtained by solving the corresponding Euler--Lagrange hypernetted chain equations. The resulting phase diagram is very similar to the one originally obtained by Ceperley and collaborators~\cite{Ceperley}, although significant differences arise at large densities. The phase diagram shows that any fermionic mixture of pure elements will always be seen in gaseous form, as the mass ratios required for crystallization of weakly bound fermionic molecules are far beyond the ones that can be achieved in nature. Finally, we investigate an alternative mechanism based on the confinement of one of the species to a deep optical lattice which exponentially increases its effective mass as a function of the confining amplitude. The  resulting mass ratio of the mixture created in this way can then be tuned at will and could be used to check experimentally the predicted phase diagram both in the gas and crystal (superlattice) phases.

\acknowledgments

The work was sponsored by (Spain) Grant No.~FIS2008-04403 and Generalitat de Catalunya Grant No.~2009SGR-1003. G.E.A. acknowledges financial support from MEC (Spain).

\end{document}